\documentclass{statsoc}
\usepackage[left=7mm,right=15mm]{geometry} 
\usepackage{vmargin,bbm,times,natbib,graphicx,amssymb}
\newtheorem{prop}{Proposition}
\newtheorem{ex}{Example}

\title[Estimating Diversity via Ratios]{Estimating Diversity via Frequency Ratios}

\author[Amy Willis]{Amy Willis}
\address{Cornell University, Ithaca, New York, U.S.A.}
\email{adw96@cornell.edu}
\author[Willis \& Bunge]{John Bunge}
\address{Cornell University, Ithaca, New York, U.S.A.}

\begin{document}

\begin{abstract}
We wish to estimate the total number of classes in a population based on sample counts, especially in the presence of high latent diversity.  Drawing on probability theory that characterizes distributions on the integers by ratios of consecutive probabilities, we construct a nonlinear regression model for the ratios of consecutive frequency counts.  This allows us to predict the unobserved count and hence  estimate the total diversity.   We believe that this is the first approach to depart from the classical mixed Poisson model in this problem.  Our method is geometrically intuitive and yields good fits to data with reasonable standard errors. It is especially well-suited to analyzing high diversity datasets derived from next-generation sequencing in microbial ecology. We demonstrate the method's performance in this context and via simulation, and  we present a dataset for which our method outperforms all competitors.

\end{abstract}


\section{Introduction} \label{intro}

Our goal is to estimate the total number of classes $C$ in a population, based on a sample of individuals from the population.  This problem has many applications in the natural sciences, as well as in linguistics and computer science, but our particular interest is in microbial ecology: estimating the biodiversity (number of taxa or species richness) in a microbial community from a sample of DNA or rRNA sequences. The rapid development of next generation sequencing technology has provided the opportunity to analyze very large microbial community composition datasets, often containing more than $10^9$ sequences. Species richness analysis is a data reduction tool that frequently provides an indication of ecosystem health \citep{li2014potential,dethlefsen2008pervasive,gao2013comparison}. However, classical approaches to the ``species problem'' perform poorly in the microbial context, because these datasets differ structurally from animal abundance datasets. In particular, microbial datasets are characterized by a large number of rarely observed species, including a high ``singleton'' count (species observed exactly once), as well as a small number of very abundant species.  The resulting large peak to the left and long tail to the right make inference especially challenging \citep{lladser, bwwa}. 

Let $f_1, f_2, \ldots$ denote the numbers of taxa observed once, twice, and so on, in the sample, and let $f_0$ denote the number of unobserved taxa, so that $C = f_0+f_1+f_2 +\ldots$.  Using the observed data we wish to predict $f_0$ and hence estimate $C$. Our approach involves analysis of the {\em frequency ratios} $f_{j+1}/f_j$, as a function of $j$.  We fit a nonlinear regression to the ratios, projecting the fitted function downward to $j=0$ so as to predict $f_0$ and estimate $C$, with associated standard errors and goodness-of-fit assessments.

The idea of using ratios of frequencies, or ratios of the probabilities $p_j$ from a probability mass function on the nonnegative integers, dates back at least to \cite{katz}, who proved that the function  $(j+1)p_{j+1}/p_j$ is linear in $j$ only for the binomial, Poisson or negative binomial distributions.  We will discuss many subsequent developments, but the introduction of the ratio plot in the species problem is due to \cite{rbbo}, who exploited the Katz structure but found that a log-transformation was needed to fit the underlying linear model to data.  Here we make a broad generalization of that work.  We fit a heteroscedastic, correlated nonlinear regression function to the (untransformed) graph of $(j,f_{j+1}/f_j)$, based on a theory of probability ratios due to \cite{kemp}.  This gives rise to a rich class of models which generate plausible estimates of $C$, as well as  standard errors and a model selection procedure.  Our method is:
\begin{itemize}
\item Geometrically intuitive, as the fits of competing curves are clear from the ratio plot;
\item More general than a linear model, and does not require transformation of the data to approximate linearity;
\item Applicable to non-mixed Poisson or to bounded-support marginal distributions;
\item Highly stable, especially compared to maximum likelihood; and
\item Sensitive with respect to identifying ``classical'' (Poisson and negative binomial) vs.\ non-classical marginal  distributions.
\end{itemize}

The traditional approach to the species problem, which dates back to \cite{fcwm} and is the basis for almost every existing method, works directly with the frequency counts.  In this setup the sample counts of the  taxa are modeled as $C$ independent Poisson random variables where the Poisson means are an i.i.d.\ sample from some mixing distribution.  The frequencies $(f_1, f_2, \ldots)$ then constitute a random sample from a zero-truncated mixed Poisson distribution, and estimation of $C$ is based the likelihood function.  This model has been studied from the frequentist, Bayesian, parametric, and nonparametric points of view (e.g., \cite{bohning06,mlin,catchall,bwwa}), and improving the stability of mixed Poisson models motivated the first ratio-based approach of  \cite{rbbo} (see also \cite{bblg,dankmarnew}).  In this paper we break away from the assumptions of the mixed Poisson model. We believe that this departure is as yet unexplored in the literature. This approach achieves greater flexibility in modeling, is underpinned by fewer assumptions, and permits simple diagnostics for model misspecification.  Our method is formal rather than exploratory ({\it viz.}, we wish to obtain a richness estimate and a standard error), so we do not contrast our method to heuristic approaches such as the ``count metameter''  of \cite{hoag}. 

We are interested, then, in models for $f_{j+1}/f_j$, or equivalently $p_{j+1}/p_j$, as a function of $j$.  The three main lines of research on this topic may be summarized as follows:
\begin{enumerate}
\item Katz' result as extended by \cite{kemp}.  Here we model the ratio plot by ratios of polynomials in $j$, which leads to a rich and flexible class of procedures that includes the Poisson and negative binomial (gamma-mixed Poisson) as special cases.  This is our primary model and we discuss it in detail below.
\item  The Lerch distribution \citep{zoal,jkk}, characterized by probability ratios of the form
\[
\frac{p_{j+1}}{p_j} = \alpha \left( 1- \frac{1}{\beta+j+1}\right)^\nu,
\]
$0<\alpha<1, \beta>0, \nu\neq0$.  This arises as the stationary distribution of a birth and death process, and so could offer a plausible model. Our investigations found that it does not generate a flexible class of statistical procedures and is difficult to fit numerically, and appears to present no advantage over the ratio-of-polynomials models.  
\item Randomly stopped sums \citep{peve}, with probability ratios
\[
\frac{p_{j+1}}{p_j} = \alpha + \frac{\beta}{\sum_{i=0}^j \gamma^i},
\]
$\alpha, \beta \in \mathbbm{R}, -1<\gamma<1$.  We have not found any relevant interpretation or advantage for these models in this problem and so do not pursue them here. 
\end{enumerate}

In summary, we define a nonlinear model, based on ratios of polynomials, for the ratios of counts.  This entails a generalized probability model for the count data which need not be mixed Poisson.  We fit the models by nonlinear regression rather than maximum likelihood, for reasons we explain below.  While the implementation is nontrivial due to the heteroscedastic and autocorrelated nature of the ratio data, the result is a flexible procedure which allows estimation of the total number of classes in a broad array of situations.  The linear model of \cite{rbbo} is encompassed as a special case. In the following sections we discuss the statistical approach, and we describe an R package called breakaway which implements the method. We analyze several datasets and present simulation results.

\section{Distributions  based on ratios of probabilities}

Let $\mathbf{p}$ denote a probability distribution on $\{0,1,2,\ldots\}$ with $p_j = P(j), j=0,1,2,\ldots$ and $\sum_j p_j = 1$.  A rich literature has examined characterization of distributions via ratios of adjacent probabilities $p_{j+1}/p_j, j=0,1,2,\ldots$.  \cite{kemp} provided the first broad theory (see also \cite{dace,kem1}), with an analysis of distributions with probability generating functions of the form 
\begin{equation} \label{hgpgf}
G(s) = \sum_{j=0}^\infty s^j p_j = \frac{_pF_q(a_1,\ldots,a_p; b_1,\ldots,b_q; \lambda s)} {_pF_q(a_1 ,\ldots, a_p; b_1,\ldots,b_q; \lambda)},
\end{equation}
where $F$ is the generalized hypergeometric function and $a_1,\ldots,a_p, b_1,\ldots,b_q$ and $\lambda$ are parameters.  Analyzing the relevant parameter spaces was one of the main points of Kemp's original paper, and we do not reproduce her results here. For these distributions the ratios of probabilities have the form
\begin{equation} \label{ratfn}
\frac{p_{j+1}}{p_j} = \frac{(a_1+j) \cdots (a_p+j)\lambda}{(b_1+j) \cdots (b_q+j)(j+1)},
\end{equation}
that is, rational functions of $j$. \cite{tg} discuss the case $a_1 = 1$.

The results of \cite{katz} clearly demonstrate that at least some distributions in the class defined by (\ref{hgpgf}), which we call Kemp-type, are also mixed Poisson. Because of the prevalence of mixed Poisson distributions in the species problem literature, it is natural to ask whether this is true of all Kemp-type distributions. The answer is negative, as we will see with the following example. 
\begin{ex} Terminating (bounded support) distributions can arise under the framework of \citet[Case (c)]{kemp}. Mixed Poisson distributions necessarily have full support. \end{ex}

Our next example suggests that  even if we restrict to Kemp distributions with full support on the nonnegative integers, these distributions need not be mixed Poisson.
\begin{ex} Suppose $p=2, q=1$, and choose $a_1 = a_2 = 1$ and $b_1 = \lambda = 0.1$, a valid choice to ensure a proper distribution \citep{kemp}. \cite{gold} prove that a probability generating function corresponds to a Poisson mixture if and 
only if $G(\cdot)$ is defined, has continuous derivatives of all order and satisfies
\[
G(1) = 1, 0\leq G(s) \leq 1, \mbox{ and } 0 \leq G^{(n)}(s)< \infty, n= 1,2,\ldots,
\]
where $G^{(n)}$ denotes the $n$th derivative of $G$, for all $-\infty < s < 1$. However, $G(-5) = -0.6365 < 0$ so the corresponding distribution is not mixed Poisson. 
\end{ex}
We conclude that the Kemp family provides an interesting direction of departure from mixed Poisson distributions. We will see in Sections \ref{sims} and \ref{data} the advantages of this generalization when analyzing microbial data.  

\section{Frequency count ratio statistics}

We begin by studying the joint distribution of the ratios $f_{j+1}/f_j, j=0,1,\ldots$.  There are $C<\infty$ classes in the population.  Assume that the $i$th class  contributes $X_i$ members to the sample, $X_i = 0, 1, \ldots$, and that $X_1, \ldots, X_C \sim$  i.i.d.\ $\mathbf{p}$, where $\mathbf{p}$  may have bounded or unbounded support.  Then $f_j = \#\{X_i = j\}$, $j=0,1,\ldots$, and define the number of observed species $c = \sum_{j\geq1} f_j$ and the number of observed individuals $n = \sum_{j\geq 1} jf_j$; $f_0$ is unobserved.  We seek expressions for the mean and covariance of $\{f_{j+1}/f_j\}_{j\geq1}$ for modeling.

Note first that the joint distribution of $f_0, f_1, \ldots$ is multinomial with (in general) unbounded support, with probability mass function (p.m.f.)
\[
\frac{C!}{\prod_{j\geq0} f_j!}\prod_{j\geq0}p_j^{f_j}.
\]
In a practical situation we can set the maximum frequency to be some fixed value $\tau$ (more on this below).  We therefore choose $\tau$ so that $\sum_{j>\tau}p_j$ is small, and we replace $\mathbf{p}$ by $\mathbf{q}\approx \mathbf{p}$ where $q_j = p_j/(1-\sum_{j>\tau}p_j), j=0,\ldots,\tau$ and $q_j=0, j>\tau$.  The p.m.f.\ of $f_0, f_1, \ldots, f_\tau$ then becomes an ordinary multinomial with bounded support,
\[
\frac{C!}{\prod_{j=0}^\tau f_j!}\prod_{j=0}^\tau q_j^{f_j}.
\]
It is not obvious how to analyze the moments of $\{f_{j+1}/f_j\}_{1\leq j \leq\tau-1}$ directly, but a Poisson approximation is available.  One version is due to \cite{mcdo}, who compared the vector $\mathbf{f} := (f_1,\ldots,f_\tau)$ with $\mathbf{V} := (V_1,\ldots,V_\tau)$, where $V_1,\ldots,V_\tau$ are independent Poisson random variables with $E(V_i) = Cq_i$.
\begin{prop} \cite{mcdo}
\[
\sum_{\mathbf{a}\in \mathbf{R}^\tau}|\mbox{Pr}(\mathbf{f}= \mathbf{a}) - \mbox{Pr}(\mathbf{V} = \mathbf{a})| \leq 2 C (1-q_0)^2 .
\]
\end{prop}
McDonald notes that this bound is useful when $q_0$ is large, although it actually must be large relative to $C$.  In the microbial ecology application $q_0$ is typically quite large, but $C$ may be so as well.  Tighter bounds are have been developed \citep{roo2,roo1}, but in too complex a form to discuss here.  We will proceed by treating $f_1,\ldots,f_\tau$ as independent Poisson random variables with $E(f_i) = Cq_i$.

We are interested, then, in the first and second (joint) moments of $\{f_{j+1}/f_j\}_{1\leq j \leq\tau-1}$.  However, with positive probability any $f_j$ may be zero.  We therefore condition on all $f_j$ being nonzero, for $j=1$ up to some maximum $J\leq \tau-1$, i.e., on the event $\{f_1>0\} \cap \cdots \cap \{f_{J+1}>0\}$.  Write $g_j := f_j|f_j>0$; by the independent Poisson assumption the $g_j$ are independent zero-truncated Poisson random variables, $j=1,\ldots,J+1$.  Let $\mu_j$ and $\sigma^2_j$ denote the mean and variance of $g_j$, respectively.

Applying the delta method, we have
\[
E\left(\frac{f_{j+1}}{f_j}\right) \approx \frac{\mu_{j+1}}{\mu_j},
\]
$j=1,\ldots,J$, and
\begin{eqnarray} \label{covf}
\nonumber \mbox{Cov}\left( \frac{f_2}{f_1}, \frac{f_3}{f_2}, \ldots, \frac{f_{J+1}}{f_J} \right)   \hspace{3.25in}
\nonumber \\ 
\approx \left[ \begin{array}{ccccc}
\frac{\mu_2^2\sigma_1^2}{\mu_1^4} + \frac{\sigma_2^2}{\mu_1^2} & &  &  &  \\
-\frac{\mu_3\sigma_2^2}{\mu_2^2\mu_1} & \frac{\mu_3^2\sigma_2^2}{\mu_2^4} + \frac{\sigma_3^2}{\mu_2^2} &  &   &  \\
0 & -\frac{\mu_4\sigma_3^2}{\mu_3^2\mu_2} & \frac{\mu_4^2\sigma_3^2}{\mu_3^4} + \frac{\sigma_4^2}{\mu_3^2} &  &   \\
\vdots &  &  \ddots & &  \\
0 & \cdots &   -\frac{\mu_J\sigma_{J-1}^2}{\mu_{J-1}^2\mu_{J-2}} & \frac{\mu_J^2\sigma_{J-1}^2}{\mu_{J-1}^4} + \frac{\sigma_J^2}{\mu_{J-1}^2} & 
\\
0 & \cdots &  0 & -\frac{\mu_{J+1}\sigma_J^2}{\mu_J^2\mu_{J-1}} & \frac{\mu_{J+1}^2\sigma_J^2}{\mu_J^4} + \frac{\sigma_{J+1}^2}{\mu_J^2}
\end{array}
\right] .
\end{eqnarray}
We regard the $f_j$ as independent zero-truncated Poisson random variables with Poisson parameter $Cq_j$, or ignoring the truncation at $\tau$, $Cp_j$.  The mean and variance of a zero-truncated Poisson random variable with (original) parameter $\lambda$ are $\lambda/(1-e^{-\lambda})$ and $(\lambda/(1-e^{-\lambda}))(1-\lambda/(e^\lambda-1))$ respectively.   Therefore $\mu_j \approx Cp_j/(1-e^{-Cp_j})$ so that 
\[
\frac{\mu_{j+1}}{\mu_j} \approx \frac{p_{j+1}}{p_j} \frac{1-e^{-Cp_j}}{1-e^{-Cp_{j+1}}},
\]
and since $C$ is typically large we regard $f_{j+1}/f_j$ as a reasonable estimate of $p_{j+1}/p_j$.  The expressions in the covariance matrix are more complicated.  We have
\begin{eqnarray}  \label{covapprox}
\mbox{Cov}\left(\frac{f_j}{f_{j-1}}, \frac{f_{j+1}}{f_j} \right) \approx - \frac{1}{C} \frac{p_{j+1}}{p_{j-1}p_j} \frac{1-e^{-Cp_{j-1}}}{1-e^{-Cp_{j+1}}} (1-e^{-Cp_j}-Cp_je^{-Cp_j}),
\end{eqnarray}
$j=2,\ldots,J$ and
\begin{eqnarray} \label{varapprox}
\mbox{Var}\left( \frac{f_{j+1}}{f_j} \right) 
\approx \frac{1}{C} \bigg{(} \frac{p_{j+1}^2}{p_j^3} \frac{(1-e^{-Cp_j})^3 }{(1-e^{-Cp_{j+1}})^2}(1-Cp_j/(e^{Cp_j}-1)) \nonumber \\
+ \frac{p_{j+1}}{p_j^2} \frac{(1-e^{-Cp_j})^2}{1-e^{-Cp_{j+1}}} (1-Cp_{j+1}/(e^{Cp_{j+1}}-1) ) \bigg{)} , 
\end{eqnarray}
$j=1,\ldots,J$.  We will return to these later when considering weights for nonlinear regression.

\section{Nonlinear regression}

Our initial model is based on (\ref{ratfn}), which we write as a ratio of polynomials in $j$.  We use nonlinear regression, since no explicit likelihood is available in general, and estimation via empirical probability generating functions also presents difficulties in this context \citep{ng}.  The standard setup then gives
\begin{equation} \label{breakaway1}
\frac{f_{j+1}}{f_j} = \frac{\beta_0^*+\beta_1^*j + \ldots + \beta_p^*j^p}{1+\alpha_1^*j + \ldots + \alpha_q^*j^q} + \epsilon_j.
\end{equation}
In order to reduce the correlation between the parameter estimates we center $j$ at $\bar{j}$, the (empirical) mean of $j$. Our final model is therefore
\begin{equation} \label{breakawayfinal}
\frac{f_{j+1}}{f_j} = \frac{\beta_0+\beta_1(j-\bar{j}) +\ldots  + \beta_p(j-\bar{j})^p}{1+\alpha_1(j-\bar{j}) + \ldots + \alpha_q(j-\bar{j})^q} + \epsilon_j,
\end{equation}
where we assume that $\mbox{Cov}([\epsilon_j])$ is given by (\ref{covf}) -- (\ref{varapprox}).
We estimate $(\beta_0,\ldots,\beta_p;\alpha_1,\ldots,\alpha_q)$ by $(\hat{\beta}_0,\ldots,\hat{\beta}_p;\hat{\alpha}_1,\ldots,\hat{\alpha}_q)$ using a preimplemented nonlinear least squares solver \citep{nls}. 

Assume for the moment that we have selected a pair $(p,q)$.  The parameter estimation problem is then
\begin{eqnarray}  \label{lsestimator}
{\arg\!\min}_{\beta_0, \ldots ,\beta_p,\alpha_1,\ldots ,\alpha_q} (\mathbf{F} - \mathbf{P})'\mathbf{W}^{-1}(\mathbf{P}) (\mathbf{F} - \mathbf{P}),
\end{eqnarray}  
where
\[
\mathbf{F} = \left[ \frac{f_{j+1}}{f_j} \right]
\]
and
\[
\mathbf{P}  = \left[  \frac{\beta_0 + \beta_1 (j-\bar{j}) + \ldots + \beta_p (j-\bar{j})^p}{ 1 + \alpha_1(j-\bar{j}) + \ldots + \alpha_q (j-\bar{j})^q}   \right],
\]
$j=1,\ldots,J$, and  $\mathbf{W}$ is the tridiagonal covariance matrix with diagonals given in (\ref{varapprox}) and off-diagonals  in  (\ref{covapprox}).  We find that numerical convergence is almost never achieved when $\mathbf{W}$ is tridiagonal, so henceforth we approximate $\mathbf{W}$ by its diagonal. Concurring with \cite{rbbo}, simulations show this results in only a slight loss of precision. 

The next question is what initial weighting scheme to use.  Prior to model selection we do not have a form for (even the diagonal) of $\mathbf{W}(\mathbf{P})$. We considered various smooth functions for this including $j^\gamma$ and $e^{\gamma j}$, and after much testing concluded that initial weights $1/j$ work well while remaining (provisionally) independent of model  selection.

To find starting values for the parameters $(\beta_1,\ldots,\beta_p;\alpha_1,\ldots,\alpha_q)$ we use a sequential procedure as in \cite{startingvalues}.   Under parametrization (\ref{breakawayfinal}) all models are nested. The parameters of the lowest order model $q=0,p=1$ can be estimated using ordinary least squares. The starting values for model  $p=q=1$ are then $\hat{\beta}_{0,OLS}$, $\hat{\beta}_{1,OLS}$ and $\alpha_1=0$. This procedure is then repeated, using each model's estimates as the initial values for the next highest order model. If a model does not converge, the starting value construction simply skips the nonconvergent model.

For $j=0$, (\ref{breakawayfinal}) gives
\[
E\left(\frac{f_1}{f_0}\right) = \frac{\sum_{r=0}^p \beta_r (-1)^r \bar{j}^r}{1 + \sum_{r=1}^q \alpha_r (-1)^r \bar{j}^r}.
\]
Rearranging and substituting parameter estimates for their unknown values, we obtain
\[
\hat{f}_0 = f_1 \left(\frac{\sum_{r=0}^p \hat{\beta}_r (-1)^r \bar{j}^r}{1 + \sum_{r=1}^q \hat{\alpha}_r (-1)^r \bar{j}^r} \right)^{-1} = \frac{f_1}{\hat{b}_0},
\]
where
\[
\hat{b}_0 := \frac{\sum_{r=0}^p \hat{\beta}_r (-1)^r \bar{j}^r}{1 + \sum_{r=1}^q \hat{\alpha}_r (-1)^r \bar{j}^r},
\]
and finally $\hat{C} = c + \hat{f}_0$.

Given the ability to computationally fit specific models, the question of model selection arises.  We select the lowest order model meeting the following conditions.  First, we require  $\hat{b}_0>0$ so that $\hat{f}_0>0$.  (The issue of negative unobserved diversity estimators is as old as the problem itself: \cite{fcwm} recognized it when fitting the negative binomial distribution to Malayan butterfly data; see also \cite{chaobunge,rbbo}).  Second, $1+\alpha_1(j-\bar{j}) + \ldots + \alpha_q(j-\bar{j})^q$ can have no roots in $[0,J]$, so that (\ref{breakawayfinal}) has no singularities in the relevant domain. Finally, the model must be computable, that is, must converge numerically. We advocate selecting the most parsimonious available model out of all models that satisfy these requirements. 

Our final procedure, implemented in the R package {\em breakaway}, is then as follows.  If a model converges numerically, yields a positive $\hat{b}_0$, and has no singularities in the relevant domain we say it ``satisfies the criteria.''
\begin{enumerate}
\item Confirm that  the data has appropriate structure, i.e., multiple contiguous frequency counts after the singleton count.
\item Fit all models as in (\ref{breakawayfinal}) with a diagonal weighting matrix and weights proportional to $1/j$.
\begin{enumerate} 
\item If no models satisfy the criteria, output the log-transformed Katz model of \cite{rbbo}. STOP.  (breakaway code 1.)
\end{enumerate}
\item Based on the fitted values of the smallest model that satisfied the criteria, recalculate the weights based on (\ref{varapprox}). 
\item Fit all models as in (\ref{breakawayfinal}) using the weights calculated in 3. Note that the weights are model independent.
\begin{enumerate}
\item Repeat 3 -- 4 until $\hat{f}_0$ stabilizes. Conclude the implied model structure. (breakaway code 2.)
\item If the first adaptively chosen weights do not result in a model that satisfies all criteria, return the smallest model based on the $1/j$ weights that did satisfy all criteria. (breakaway code 3.)
\end{enumerate}
\end{enumerate}

We now discuss error estimation. We apply the delta method to find an estimate of $\mbox{Var}(\hat{f}_0)$:
\begin{eqnarray} 
\hat{\mbox{Var}}(\hat{f}_0) \approx f_1\hat{b}_0^{-2} \left( 1-\frac{f_1}{n} + f_1 \hat{b}_0^{-2}\hat{\mbox{Var}}(\hat{b}_0)\right),
\end{eqnarray}
where $\hat{\mbox{Var}}(\hat{b}_0)$ is an empirical estimate of 
\begin{eqnarray*}
& \left(\nabla \left(\frac{\sum_{r=0}^p \beta_r (-1)^r \bar{j}^r}{1 + \sum_{r=1}^q \alpha_r (-1)^r \bar{j}^r}\right)\right)^T  \times \mbox{Cov}(\hat{\beta}_0,\ldots,\hat{\beta}_p,\hat{\alpha}_1,\ldots\hat{\alpha}_q) \times \nabla \left(\frac{\sum_{r=0}^p \beta_r (-1)^r \bar{j}^r}{1 + \sum_{r=1}^q \alpha_r (-1)^r \bar{j}^r}\right),
\end{eqnarray*}
and we treat the covariance between $\hat{b}_0$ and $f_1$ as negligible. This yields 
\begin{eqnarray} \label{finalvar}
\hat{\mbox{Var}}(\hat{C}) \approx \frac{n \hat{f}_0}{n+\hat{f}_0} + \hat{\mbox{Var}}(\hat{f}_0),
\end{eqnarray}
again treating the covariance term as negligible (as in \cite{rbbo}). A simulation study in the negative binomial case (Table \ref{t:errors}) shows this approximation to be remarkably accurate. 

Upon carrying out our estimation several outcomes are possible.  First, the estimates of the parameters $\alpha_i$ and $\beta_i$ may fall within the parameter spaces defined by \cite{kemp}, or even within the Katz subset of the Kemp-type distributions, namely the negative binomial or Poisson.  In this case we presume that the data is well-described by such a distribution; below we show a simulation of the negative binomial case indicating good performance and identification of that distribution.  Second, the estimates may imply a legitimate terminating Kemp-type distribution.  Third, the parameter estimates may not entail a distribution at all (the implied $p_j$ would be negative); in this situation we can still use our method to estimate $C$ but cannot give an interpretation to the probability model.  Below we see that this behavior occurs on the Epstein dataset.

\section{Simulations} \label{sims}

Analyzing breakaway's behaviour for simulated negative binomial frequency counts demonstrates that when the true distribution is negative binomial, breakaway correctly infers this in more than 99\% of cases, see Table \ref{t:errors}. Note that Table \ref{t:errors} is based on simulating negative binomial counts near the boundary of the parameter space, and hence breakaway performs well even in a near-pathological case. These choices of parameters lead to relatively high frequencies of rare species (singletons, doubletons, etc.), which is consistent with data structures observed in microbial diversity studies. 

\begin{table}
\caption{\label{t:errors}   Mean estimated standard error (calculated using (\ref{finalvar})) and actual standard error in estimating $\hat{C}$ when the true frequency count distribution is negative binomially distributed. The percentage of replications that resulted in breakaway correctly inferring the negative binomial (NB) distribution is also shown. The true value of $C$ and true probability and size parameters are given. Results are based on 10,000 replications.} \\ 
\begin{tabular}{ l | l | c |  c |c} \hline       
  True $C$ & (probability, size) & \% inferred NB & $\hat{\mbox{s.e.}}(\hat{C})$ & True s.e.($\hat{C}$) \\ \hline 
  50,000 & $(0.99,500)$ &99.57 & 20.69 & 20.84 \\ 
  50,000 & $(0.95,100)$  & 100.00 & 19.56 & 19.73  \\
  5,000 & $(0.99,500) $  & 99.70 & 6.61 & 6.57 \\
  5,000 & $(0.95,100)$  & 99.99 & 6.24 & 6.20 \\
  \hline  
\end{tabular}
\end{table}


\section{Applied data analysis} \label{data}

We demonstrate our approach on three microbial datasets, entitled Apples \citep{apples}, Soil \citep{Ursel}, and Epstein (S. Epstein, personal communication, February 28, 2014), all available as supplementary materials. We consider Apples a medium-diversity case, and Soil and Epstein as high-diversity.

Table \ref{t:estimates} compares the results of breakaway with a number of competitors: the transformed and untransformed weighted linear regression model (tWLRM and uWLRM) of \cite{rbbo}, the Chao-Bunge estimator \citep{chaobunge}, the \cite{chao} lower bound (CLB), and 95\% bootstrap confidence intervals for the nonparametric maximum likelihood (NPMLE) estimator of \cite{nopo} (c.f.\ also \cite{wali,rspecies}). Following standard practice we set $\tau=\tau_{max}$ for the WLRMs (where $\tau_{max}$ is the largest frequency before the first zero frequency) and $\tau=10$ for the less robust Chao-Bunge estimator. breakaway employs $\tau_{max}$. The diagonal weighting structure for breakaway was employed as the tridiagonal model does not converge in any of these cases. 

\begin{table}
\caption{\label{t:estimates}  Comparison of the method breakaway with other diversity estimators: the transformed and untransformed weighted linear regression model (tWLRM and uWLRM), Chao-Bunge estimator, Chao lower bound (CLB), and 95\% bootstrap confidence intervals (1000 resamples) for the nonparametric maximum likelihood estimator (NPMLE). Standard errors are given in parentheses where appropriate.}
\begin{tabular}{ l | l | l | l } \hline       
  Dataset & Apples & Soil  & Epstein \\ \hline 
  breakaway &  1552 (305)  & 5008 (689)& 2162 (1699) \\ 
	uWLRM&  1330 (77) &7028 (1743) & *  \\
	tWLRM & 1179 (28) & 3438 (87) & 1849 (380) \\
	Chao-Bunge & 1377 (63) & 4865 (257) &  *  \\
	CLB & 1241 (38) & 3674 (84)  & 1347 (160) \\
	NPMLE & (1049,2776)&  (2885,38442) &(730,6936721) \\
  \hline  
\end{tabular}
\end{table}

We observe that in the medium diversity setting (Apples), breakaway's estimate is comparable to other estimators after noting its larger standard error. However, the  advantage of breakaway is revealed in the high diversity cases of Soil and Epstein.  breakaway's standard error is less than half that of the uWLRM in the case of Soil. In the case of the Epstein dataset, the uWLRM and Chao-Bunge estimators both fail to produce any estimate. While breakaway's standard error is large relative to the estimate, we argue that the failure of both Chao-Bunge and the uWLRM to produce any estimate suggests that a high standard error is an honest assessment of the nature of the dataset. We cite the instability of the NPMLE as further evidence for this claim.

\begin{center}
\begin{figure}
\includegraphics[trim= 0mm 0 0 0, scale=0.7]{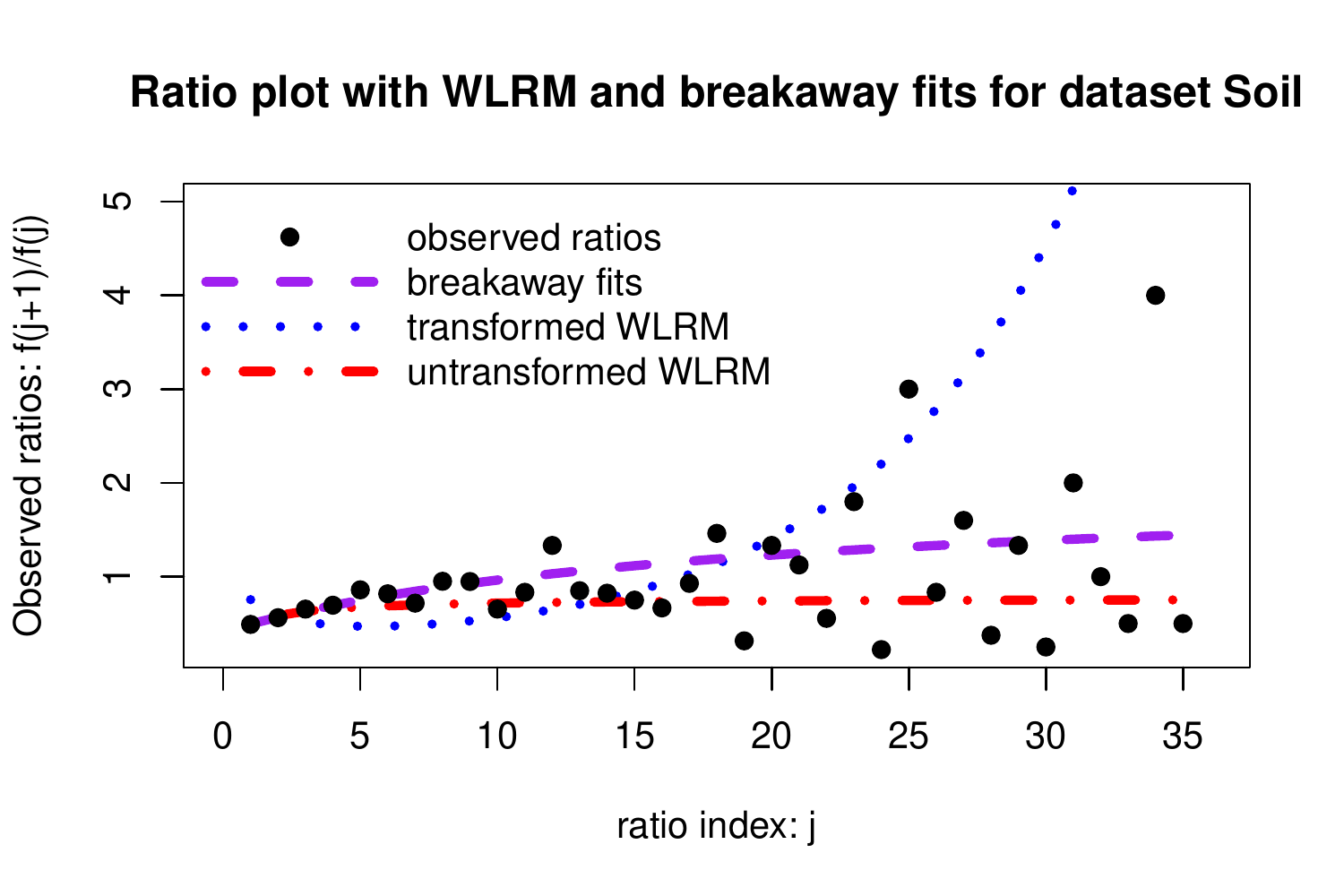}
\caption{Comparison of fitted ratios for regression-based diversity estimators: the transformed and untransformed weighted linear regression model (WLRM), and breakaway. The above pertains to the Soil dataset.}
\label{p:fits2}  
\end{figure}
\end{center}

One appealing feature of regression-based estimators is that fit can be readily assessed, and the plausibility of the models employed by breakaway can be seen in Figure \ref{p:fits2} in the case of the Soil dataset (ratio plots for the other datasets appear similar). This plot elucidates that the tWLRM standard errors in Table \ref{t:estimates} are artificially low due to model misspecification. We argue that the added flexibility afforded by (\ref{breakawayfinal}) replaces the need to log-transform to achieve a positive prediction for $f_0$. 

Table \ref{t:weights} displays the empirical weights of the regression-based models for the Soil dataset. We observe that breakaway's weights, which are data-adaptive rather than based on the negative binomial model, appear to be a ``middle ground'' between the weights of the tWLRM and the uWLRM. We note that the uWLRM places more than 98\% weight on the first ratio, providing an explanation for its poor fit as displayed in Figure \ref{p:fits2}.  

\begin{table}
\caption{\label{t:weights} Comparison of empirical weights placed on the first through fifth data points for the regression-based diversity estimators: the transformed and untransformed weighted linear regression model (WLRM), and breakaway. The below pertains to the Soil dataset. }
\begin{tabular}{ l | l | c |  c | c | c} \hline       
  Model & $w_1$ & $w_2$ &$w_3$ &$w_4$ & $w_5$ \\ \hline 
Transformed WLRM &0.803 &0.431 &0.266& 0.181& 0.142 \\
Untransformed WLRM&0.983 &0.178& 0.046& 0.018& 0.006 \\
breakaway &0.859 &0.431& 0.226& 0.124& 0.071 \\
  \hline  
\end{tabular}
\end{table}


In the Soil, Apples, and Epstein datasets, the model with $p=q=1$ was found to be sufficiently flexible to produce positive predictions for $f_0$, though other studied datasets require higher order models. Parameter estimates for the Soil and Apples datasets do not correspond to distributional models with unbounded support, however if the support is considered bounded then normalization is possible and proper distributional models are implied. However, the parameter estimates for the Epstein dataset are not normalizable since the implied ``probability structure'' yields negative probabilities. We discuss the implications of this in Section \ref{conc}. That breakaway correctly infers negative binomial distributions but rarely selects them when faced with real data is suggestive. 

Similar to the uWLRM, NPML and Chao-Bunge estimators, breakaway does not guarantee a positive estimate. However, breakaway produces positive estimates in many situations where these estimators fail, and performs especially well in the high diversity situations that are typical of modern microbial datasets. The additional parameters in the ratio model contribute the flexibility required to produce positive estimates in these difficult cases and improve model fit. Furthermore, the weighting scheme is adaptive and thus more realistic than that of other regression-based estimates while preserving the appealing feature of a visual mechanism to confirm model specification. breakaway imposes no {\it a priori} distributional model, and is not constrained to the mixed Poisson framework. Estimates and standard errors are consistent with other estimators in datasets displaying low and medium diversity characteristics, and estimates produced by breakaway tend to exceed the Chao lower bound. Finally, the program is implemented in an R package freely available to the practitioner. 

\section{Conclusions and future directions}\label{conc}


We have developed a diversity estimator based on fitting a class of distributions characterized by the form of their probability ratios. Distribution theory tells us that this form is nonlinear, heteroscedastic and autocorrelated, which we have taken into account in developing our method. The method produces reasonable estimates with sensible standard errors in low, medium and high diversity situations, and can outperform competitors in the high diversity situations that characterize modern microbial datasets. Our estimator tends to produce similar results to nonparametric estimators when they exist. Diversity estimates and standard errors are consistent with simulation studies and the fitted structures are shown to be geometrically and intuitively plausible. 

While this paper has focused on fitting ratios-of-polynomials models  to frequency ratios via nonlinear regression, other probabilistically-based functional forms of the frequency ratios could also be investigated. The Lerch distribution, which arises as the stationary distribution of a stochastic process, can be characterized by either the form of its frequency ratios or by its probability mass function, which exists in closed form (unlike general Kemp-type distributions). We attempted to estimate the parameters of a zero-truncated Lerch distribution via maximum likelihood and found that the algorithm did not converge for any of the datasets investigated above, which is a well-documented issue with direct estimation of parameters in the species problem \citep{93review, bwwa}. Furthermore, only for the Apples dataset did nonlinear regression estimates of the Lerch parameters converge, suggesting that the Lerch distribution may not be flexible enough to model a broad range of diversity datasets. In the case of the Apples dataset, the Lerch nonlinear regression fits were similar to breakaway. However, the Lerch model is unstable, estimating $\hat{C} = 1727$ $(1987)$ compared to breakaway's $\hat{C} = 1552$ $(305)$. 

Our method amounts to modeling the frequency count ratios via a  low-dimensional functional representation. By parsimoniously fitting this low-dimensional structure, the method may depart from a distribution-based structure altogether, and we have presented a dataset for which estimation is only possible if this is permitted. We argue that this does not interfere with the ultimate goal of estimating unobserved diversity and that inference remains valid. However, in many contexts a distributional model is implied if the support of the frequency count distribution can be regarded as bounded. Indeed, the right extremity of the population frequency count distribution is the frequency of the most common taxon, which must be finite if the population size is believed to be finite. 

Under our procedure, models may be excluded for reasons other than goodness of fit, and classical model selection diagnostics do not apply without accounting for the conditioning inherent in this procedure. Furthermore, asymptotic properties such as consistency and normality have not been considered due to the complexity arising from the iterative nature of the algorithm. However, simulations in the negative binomial case support hypotheses of both consistency and normality. 

Finally, the sampling procedure used to construct frequency count tables in the microbial setting is complex, and many bioinformatic preprocessing tools distort and bias frequency tables \citep{qiime,uparse}. Essentially, measurement error is extreme, especially in the singleton count. Regression-based ratio methods are in an ideal position to adjust diversity estimates to account for this measurement error. This is an ongoing topic of the authors' research.


\bibliographystyle{rss}
\bibliography{EDvFR}

\end{document}